\documentclass[16pt,a4]{article}
\usepackage{epsfig}
\usepackage{setspace}
\title{Influence of Grain Size on the Band-gap of Annealed SnS Thin Films}

\author{Priyal Jain$^{a,b}$ and P.
Arun$^b$\footnote{email:arunp92@physics.du.ac.in, Telephone:091 011
29258401, Fax: 091 011 27666220} \\ \\
$^a$Department of Electronic Science,\\
University of Delhi, South Campus\\
Delhi 110021\\ \\
$^b$Material Science Research Lab,\\ S.G.T.B. Khalsa College,\\
University of Delhi, Delhi - 110 007, India\\
}

\begin{document}
\maketitle

\begin{abstract}

The manuscript reports the variation in optical band-gap of vacuum annealed 
SnS thin films. The samples were characterized by using X-Ray Diffraction,
Scanning Electron Microscopy (SEM), UV-visible Spectroscopy and Raman Analysis. 
Results show that while annealing does not effect the nano-crystalline 
sample's lattice structure or unit cell size, it does control the grain size. 
The band-gap (${\rm E_g}$) decreases with increase in grain size. ${\rm E_g}$ 
values were found to be very high (1.8-2.1~eV) for samples studied.

\end{abstract}

\vskip 2cm
\vfil \eject

\section{Introduction}
Tin sulphide (SnS) is a semiconducting material formed from elements
belonging to the IV-VI group of the periodic table. Research on SnS shows 
that it has potential use as holographic or optical data storage medium 
\cite{radot, patil}. Also, due to its high absorption 
coefficient (of the order $\sim 10^4~cm^{-1}$) and bandgap $\sim 1.3-1.6~eV$, 
SnS films have been used as photovoltaic devices \cite{zhi,noguchi}. SnS 
exists in various crystallized states like orthorhombic \cite{sohila}, 
tetrahedral (Zinc blende like) \cite{shen} or a highly distorted rock-salt 
(NaCl) structure \cite{mariano}. Due to the nature of the tin and sulphur 
bondings, it forms two dimensional sheets \cite{jiang}, giving rise to a 
layered structure with strong intra-planar forces and weak Van der Waal 
forces between the adjacent planes \cite{nikol,albanesi}. Films of SnS have 
been fabricated using various techniques such as thermal evaporation 
\cite{nahass}, RF sputtering \cite{hartman}, chemical vapour deposition 
\cite{nair}, electrodeposition \cite{ghaza}, spray pyrolysis \cite{thanga} 
and sulphurisation of metallic 
precursors \cite{reddy}. The optical properties of SnS strongly 
depend on the deposition method used \cite{gao}-\cite{ortiz}. While a maximum 
band gap (1.78~eV) was reported in films obtained by wet chemical method 
\cite{sohila}, the lowest band-gap (1.12~eV) was reported for films made by 
chemical bath deposition \cite{gao}. 

The deposition techniques influences the film's grain size, crystal 
structure and residual strain which in turn effect the optical 
properties of the film. In light of the increasing importance of SnS and its 
potential applications, it becomes important to study the film structure 
with its band-gap. Thus, in this manuscript we report our studies on vacuum 
evaporated SnS thin films that were vacuum annealed after fabrication and try 
to establish a relation between their 
structural and optical properties.

\section{Experimental}
SnS films of varying thicknesses were fabricated by thermally evaporating 
powdered SnS at vacuum better than ${\rm \sim 4x10^{-5}}$~Torr in a Hind High 
Vac (12A4D) thermal evaporation coating unit. The depositions were made on 
glass microscopy slides maintained at room temperature. SnS powder of 99\% 
purity supplied by Himedia(Mumbai) was used as the starting material. The 
thickness of the asgrown films were measured by a Veeco Dektak Surface     
Profiler (150). The samples were then subjected to post-annealing 
temperatures under vacuum for 30~mins at 373~K and 473~K. The structural, 
morphological, compositional and optical characterisations were done 
systematically for both the as grown and vacuum annealed films. We report and 
compare here the structural, morphological, compositional and optical
properties of annealed samples of five different thicknesses, 270, 480, 600,
650 and 900~nm. The stuctural analysis were done using 
Bruker D8 Diffractometer. The Raman spectra of the films were
obtained with Renishaw Invia Raman Microscope using Argon laser beam 
in the reflection mode. The surface morphology of the samples and their
chemical composition were examined with FEG-SEM JSM7600F and its
Energy-dispersive X-Ray Spectroscopy (EDX) attachment. The 
optical properties of the samples were carried out using Systronics
UV-VIS Double beam Spectrophotometer (2202).

\section{Results and Discussion}
The Raman spectra were collected from back scattered ${\rm Ar^{2+}}$ laser.
Analysis of all our samples were done maintaining constant beam
area, exposure time (30~sec) and power (15~mW). All Raman spectra were
collected using low power laser since studies on SnS as an optical data 
storage medium have shown changes induced in it by photo-thermal absorption 
of light. All the samples without
exception showed three peaks, two prominent peaks around ${\rm 170~cm^{-1}}$
and ${\rm 238~cm^{-1}}$ and a third peak of lower intensity at ${\rm
330~cm^{-1}}$. The 
${\rm 330~cm^{-1}}$ peak is associated with ${\rm SnS_2}$ \cite{Lucovsky}.
The ${\rm \sim 170}$ and ${\rm 238~cm^{-1}}$ Raman peaks are associated
with those of single crystal SnS \cite{sohila}. Our peak positions are 
slightly displaced from these
positions. Figure~(1) shows representative Raman spectra of a 900nm thick 
film, before annealing (rt) and after annealing at ${\rm 373~K}$ and ${\rm
473~K}$ respectively. The relative contribution of the ${\rm 330~cm^{-1}}$ 
peak in annealed samples is lower as compared to that in as~grown samples. 
Annealing, hence, removes ${\rm SnS_2}$ bonding/ defects formed during film 
fabrication. The ${\rm 170~cm^{-1}}$ or the ${\rm B_{2g}}$ mode's peak 
corresponds to interaction along the inter-layer `b' axis, 
while the ${\rm 238~cm^{-1}}$ peak is the ${\rm A_g}$ mode that corresponds 
to the symmetric Sn-S bonding stretching mode in the a-c plane \cite{Chandrasekhar}. 
It is clear from figure~(1) that the relative contributions from 
${\rm B_{2g}}$ and ${\rm A_g}$ do not change on annealing. 

Figure~(2) exhibits the X-Ray diffractograms of thin annealed 
(${\rm 473~K}$) SnS films of different thicknesses. Three prominent peaks 
are seen at ${\rm 2\theta \sim 31^o}$, ${\rm 37^o}$ and ${\rm 43^o}$.
These peak positions matched with those reported in ASTM Card No. 83-1758.
The SnS in our samples hence exists with orthorhombic unit cell structure 
with the (040) peak having the maximum intensity. While there are many works
in the literature that show (040) peak to be the most intense \cite{yue,Feng},
others have reported (111) peak to be the most intense \cite{sohila,devika3}. 
This again
underlines how growth technique and conditions influence the character
of the SnS films. The Miller indices of the remaining prominent peaks are 
also indicated in the figure. The lattice parameters `a', `b' and `c' of SnS in 
single crystal state are given as 4.148, 11.48 and 4.177~\AA, respectively.

We have evaluated the lattice parameters from the (040), (200) and (131) 
peak's position using the relation
\begin{eqnarray}
sin^2\theta_i=Ah^2_i+Bk^2_i+Cl^2_i\nonumber
\end{eqnarray} 
The lattice parameters of the annealed samples were found to be ${\rm 4.07
\pm 0.002}$, ${\rm 11.36 \pm 0.01}$ and ${\rm 4.45 \pm 0.002 \AA}$. They were 
found to be different from those of SnS in single crystal 
state. While the variation in `a' and `b' maybe be considered marginal
(decrease), there 
is a significant increase in `c'. Albanesi et al \cite{albanesi} have done 
systematic analyses 
to understand the effect of changing unit cell size on SnS band-gap. Their 
study suggests that an increase in volume of the unit cell would lead to a lowering 
in SnS band-gap. We shall return to this when we discuss the results of our 
optical studies. While there is a significant change in the lattice constants 
(namely `c') as compared to that of single crystal, we find that they do not 
change with film thickness or annealing
temperature. Figure~(3A) shows all three lattice parameters remaining 
uneffected with film thickness. Since points coincide, each point in this 
graph represents samples annealed at 373~K and 473~K.  
The difference in lattice constant present (${\rm c_{obs}}$) with respect to
that given in the ASTM (${\rm c_{ASTM}}$) implies that the grains are in a
state of stress with the peaks shifting left compared to those reported in
the ASTM, we conclude tensile residual stresses are acting within the grains
\cite{pat}. However, ${\rm c_{obs}}$ is constant for all the samples hence the
same tensile residual stress exists in all the samples. This tensile
residual stress thus appears as a background in our study. While the lattice 
parameters remain unaltered, the grain size changes with 
annealing. The grain size reported were calculated using the Full Width at
Half Maxima (FWHM) of the diffraction peaks in Scherrer's formula 
\cite{Cullity}. From figure~(3B), it is clear that there is a 
presistent increase in grain size with annealing, however this effect is 
more pronounced in thinner films. The grain size increases from 12~nm 
to 25~nm in the 270~nm films, while this increase in grain size with 
annealing temperature is impeded for thicker films (fig~3B). 

SnS is a layered compound with zig-zagged molecules of SnS forming layers.
These layers stack one on top of the other along the `b' axis with weak 
Van~der~Waal `inter-planar' forces acting between the layers. The layers
themselves because of the zig-zagging have finite thickness and along the
`b' axis have strong `intra-planar' forces acting within them. Ehm \cite{Ehm} 
has shown that the application of pressure decreases inter-planar distances along
`b' axis without disturbing intra-planar spacing. As stated above, the Raman
${\rm B_{2g}}$ peak represents vibrations along the `b' axis. Fig~(4) shows 
the variation of Raman peak position of annealed samples.
While the ${\rm A_{g}}$ peak remains fixed at ${\rm 226.5~cm^{-1}}$, the 
${\rm B_{2g}}$ peak shows a systematic shift to a higher wavenumber with 
increasing film thickness. It would appear thinner films with larger grain 
size have vibrations taking place at lower wave-number than thicker films 
with smaller grains. The Raman peak's position and its FWHM is a rich source
of information. While it does give information on the structure, the
corroborative  X-ray diffraction study showing no variation in lattice
parameters imply variation in Raman spectra is indicative of something else.
Literature on Raman Analysis have shown variation of FWHM to be a measure of 
phonon confinement \cite{kuninori,hyunchul} with shift in peak position 
related to grain size \cite{hyunchul}-\cite{s11} and nature of defect on the 
grain boundaries \cite{kuninori}. A shift to higher wavenumbers is indicative 
of smaller grain size. Hence, the results of X-Ray diffraction and Raman 
Analysis, when put together, suggest that annealed SnS thin films have large
grains with grain size decreasing as the annealed film thickness increases. 
This is best understood from fig~(5).

The results shown in figure~(3B) are also reflected in the 
Scanning Electron Microscope's micrographs (fig~6).
The film of 480~nm thickness showed significant increase in grain size and 
grain density when annealed at higher temperature as compared to the 
900~nm thick film. Comparing the 900~nm thick films annealed at 373~K 
and 473~K respectively, we do not find such improvement in grain
size or density as was infered from the X-Ray diffraction analysis.
Fig~(7) shows EDX of 900~nm thick film annealed at 473~K. Even after
annealing at 473~K no oxidation is evident (Si and O is from glass
substrate). This is evidently due to fact annealing was done in vacuum
(${\rm 5\times 10^{-2}}$~Torr). Based on the structural and morphological 
results, it would be 
interesting to study the optical properties of these samples. Since the 
lattice parameters and residual tensile stress on the grain's bulk remains 
same with only grain size and to an extent grain density varying 
in the samples, the study should reveal the contribution that grain size
have on
the properties like band gap, ${\rm E_g}$.

The optical absorption and transmittance spectra of all the samples were 
obtained in between wavelength range 300-900~nm. The absorption spectras were
typical with absorption increasing near band-edge. Tin sulphide is known to
have both direct band-gap \cite{Devika} and indirect band-gap \cite{sohila}. 
The band-gap can be
evaluated using the absorption coefficient (${\rm \alpha}$) obtained from the 
UV-visible spectra. The band-gap is obtained using the relation \cite{nakata}
\begin{eqnarray}
\alpha h\nu = (h\nu-E_g)^n\nonumber
\end{eqnarray}
where `n=1/2' for direct allowed transitions. On extending the best
fit line on the linear part of the curve between ${\rm (\alpha h\nu)^2}$ and 
${\rm h\nu}$, the point where the line cuts the `X-axis' gives the
energy band-gap of the sample. Using this method, we have evaluated band-gaps 
for all the samples. As 
discussed earlier, the lattice parameters remained unchanged for annealed 
films, hence it can be conclusively said that the bandgap variation is not 
caused by the lattice defects. 
The bandgap of SnS films strongly depends on the depostition process and the
film parameters like the film thickness. SnS amorphous film of thickness 
1000~nm grown on a glass substrate by spray pyrolysis has a bandgap of 1~eV
\cite{thanga}. While films grown by vacuum evaporation were found to be 
nano-crystalline with bandgap varying between 1.15-1.3~eV. The variation in 
bandgap was found to be related to the distance between the source
and the substrate \cite{devika3}. Films grown on ITO substrates by pulsed 
electrodeposition were polycrsytalline in nature with bandgap being 1.34~eV
\cite{yue}. Polycrystalline films grown by chemical bath deposition on glass 
substrates of 290~nm thickness also possess low bandgap (1.15~eV) \cite{gao}. 
While the films grown with RF sputtering technique are also nano-crystalline,
they have very high bandgaps varying from 1.3 to 1.7~eV \cite{hartman}. A 
recent study \cite{tsf} on SnS polycrystalline thin films grown by thermal 
evaporation reports SnS band-gap between 2.15-2.28~eV. Clearly,
polycrystalline SnS films have higher band-gaps compared to their amorphous
counterparts.

Figure~(8) shows the grain size dependence of the bandgap. The bandgap 
decreases with the increasing grain size. Table~1 summarizes the findings of 
this study. It shows on annealing, thinner films have larger grains. The large 
grain size gives the annealed samples a lower band-gap. Thicker films 
however, have smaller grains and an enhanced band-gap. The bandgap of 
a sample is broadly effected by its chemical 
composition, crystal structure, grain size and defects. Of all these 
contributing parameters, except for grain size, all other parameters 
remain constant. While the larger volume of the unit cell leads to a
background low ${\rm E_g}$ (${\rm \sim1.8~eV}$) as explained by Albanesi 
\cite{albanesi}, decreasing grain size contribute to an increase in 
${\rm E_g}$ of the vacuum annealed SnS samples. The band-gap's variation
with grain size due to quantum confinement has the quantitative form
\cite{eg}
\begin{eqnarray}
E_g^{nano}=E_g^{bulk}+{\hbar^2\pi^2 \over 2Mr^2}\label{eeg}
\end{eqnarray}
where `r' is the radius of the nanoparticle and `M' the effective mass of the
system. The solid line of fig~(8) shows good fit of eqn(\ref{eeg}) to our
experimental data points. The result again reiterates influence of both the
lattice parameters (via ${\rm E_g^{bulk}}$) and grain size on the band-gap of
SnS.

\section*{Conclusion}
Thin films of SnS were fabricated by thermal evaporation on microscopy glass
slides at room temperature. These films when annealed in vacuum at 373~K and
473~K developed a tensile strain along the lattice `ac' plane. The magnitude 
of this tensile strain was independent of the film thickness and annealing
temperature. However, the film thickness and annealing temperature play 
an important influence on the samples grain size. We conclude from
our results that ${\rm E_g}$ of nano-crystalline SnS films invariably depends 
on the grain size due to quantum confinement.

\section*{Acknowledgment}
Authors are thankful to the Department of Science and Technology for funding
this work under research project SR/NM/NS-28/2010. We are also grateful to
Dr Chhaya Ravi Kant, Department of Applied Sciences, IGIT (GGS-IP
University, Delhi) for extending necessary assistance.

\newpage
\section{Tables}
\begin{center}
\vskip 0.5cm
{\bf Table 1:} {\sl A comparison of state in which a very thin and thick SnS 
films exists in.}\\
\vskip 0.2cm

\begin{tabular}{c c c}
\hline 
  & Thin film & Thicker film \\ \hline
  Grain Size  & Large  &  Small \\ 
  ${\rm E_g}$  & Lower  &  Larger \\ \hline
\end{tabular}
\end{center}

\newpage
\section*{Figures Captions}
\begin{itemize}
\item[1.] A representative Raman spectra of a 900~nm thick SnS film
before and after annealing at termperatures indicated.
\item[2.] X-Ray Diffraction patterns of SnS films of thicknesses (a)
270, (b) 480, (c) 600, (d) 630 and (e) 900~nm after annealing at 373~K.
\item[3.] Graphs show the variation of (A) lattice constants with film
thickness (annealed) films. Diameter of representing points are of the
order of error in the determination of lattice parameters. (B) Grain size 
with annealing temperature. Bars indicate the ${\rm \pm1.5~nm}$ error in
determination of grain size. These results were obtained from the 
X-Ray Diffraction Patterns.
\item[4.] The variation in Raman ${\rm A_g}$ and ${\rm B_{2g}}$ peak 
positions of annealed (373~K and 473~K) SnS films with film thickness.
\item[5.] The ${\rm B_{2g}}$ peak position of Raman spectra varies with
grain size. The graph shows shift of Raman peak to lower energy level with
increasing grain size.
\item[6.] SEM micrographs show grain size improvement in thinner film
(480~nm) when annealed at higher temperature. However, the 900~nm films show
no variation in grain size on annealing.
\item[7.] Energy Dispersive X-Ray (EDX) Spectroscopy of a larger area of
900~nm thick film, vacuum annealed at 473~K, show no oxidation of the
sample. The minimal Silicon and Oxygen present are from the substrate.
\item[8.] The band-gap of annealed SnS films were found to decrease with
increasing grain size. The curve fit line agrees with theoretical model
that explains variation of band-gap with grain size.
\end{itemize}

\newpage
\section*{Figures}
\begin{figure}[h!]
\begin{center}
\epsfig{file=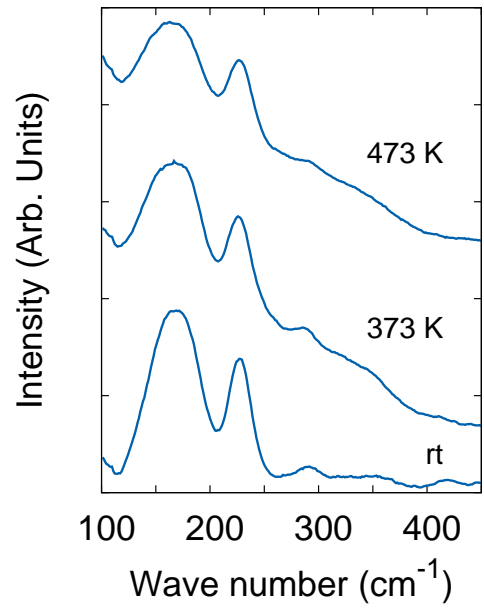, width=3.4in, angle=-90}
\end{center}
\label{raman}
\caption{\sl A representative Raman spectra of a 900~nm thick SnS film
before and after annealing at termperatures indicated.}
\end{figure}
\newpage
\begin{figure}[h!]
\begin{center}
\epsfig{file=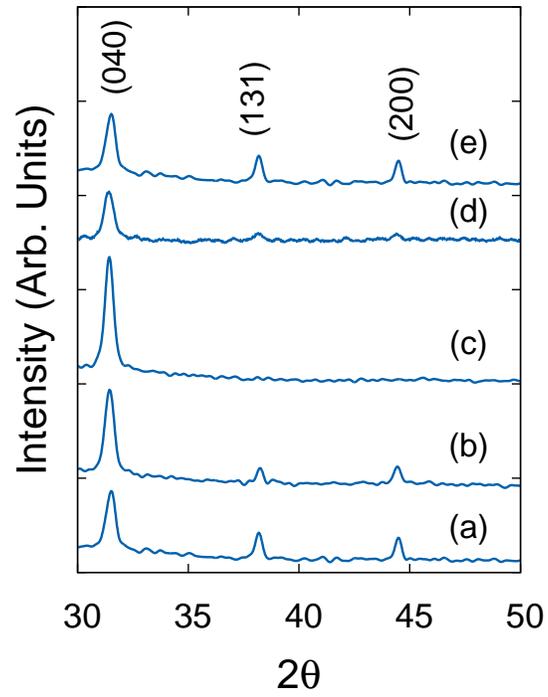, width=3.75in, angle=-90}
\end{center}
\label{fig:xrd}
\caption{\sl X-Ray Diffraction patterns of SnS films of thicknesses (a)
270, (b) 480, (c) 600, (d) 630 and (e) 900~nm after annealing at 373~K.}
\end{figure}
\newpage
\begin{figure}[h!]
\begin{center}
\epsfig{file=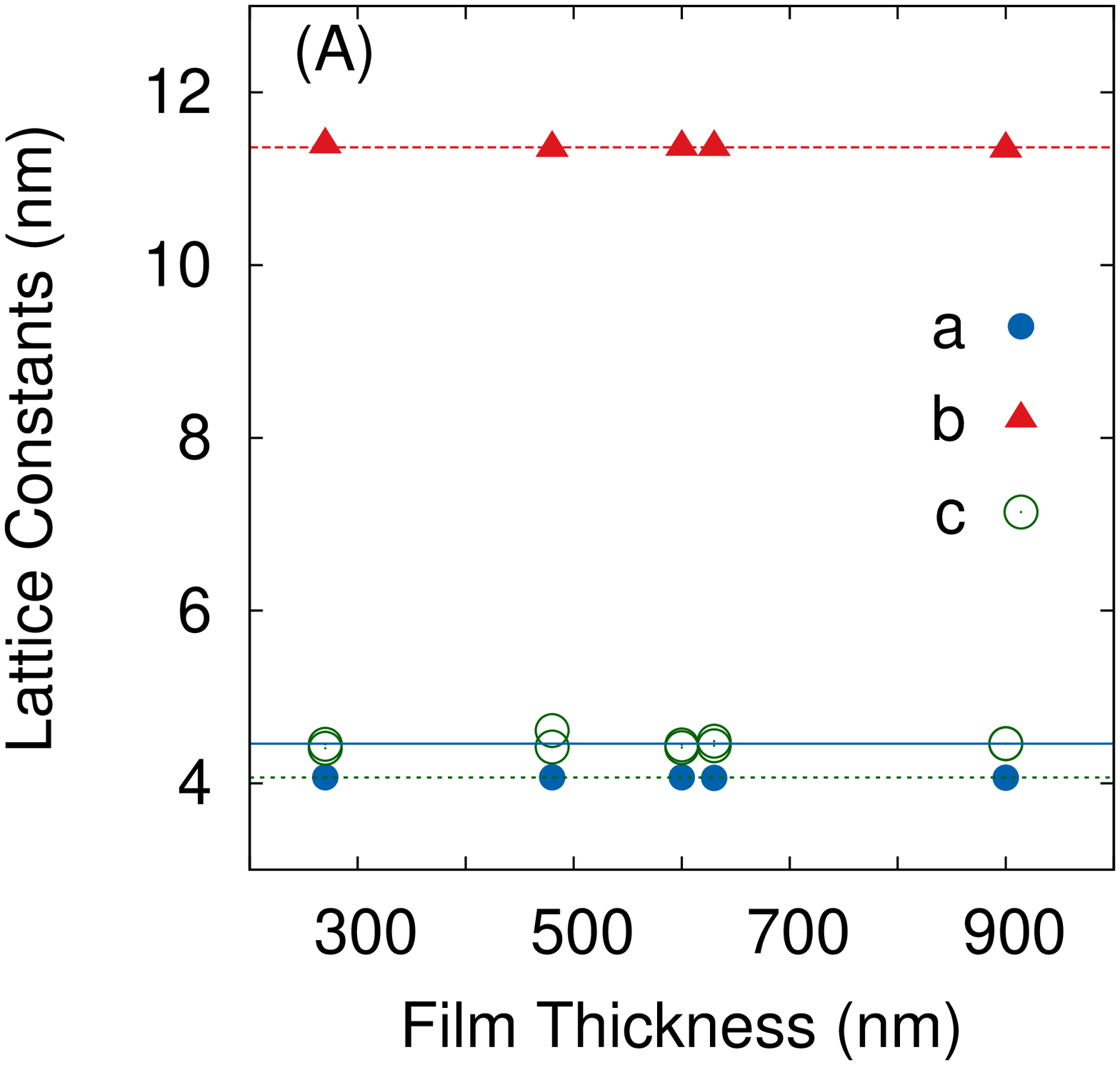, width=2.75in, angle=-90}
\hfil
\epsfig{file=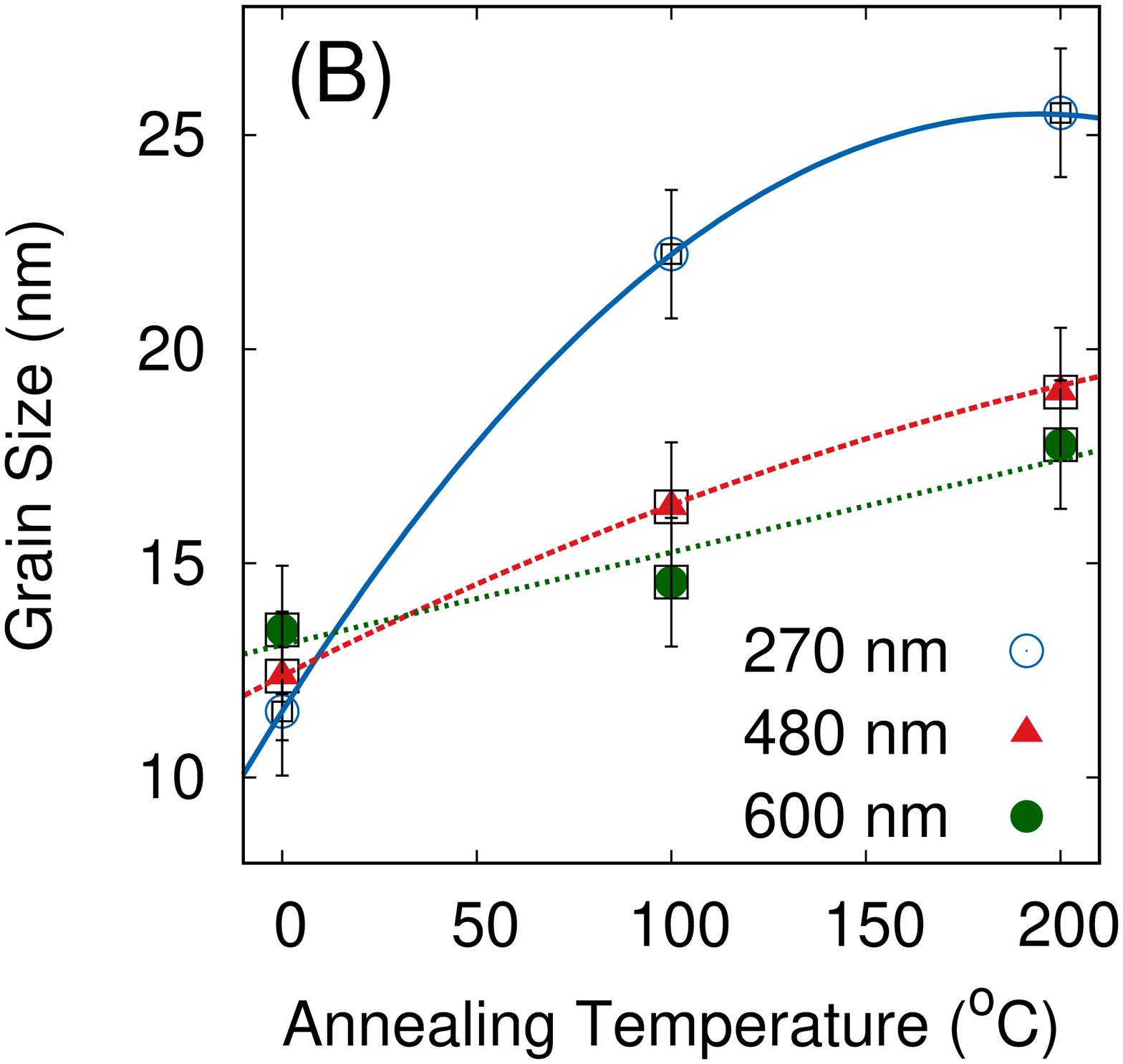, width=2.7in, angle=-90}
\end{center}
\label{res1}
\caption{\sl Graphs show the variation of (A) lattice constants with film
thickness (annealed) films. Diameter of representing points are of the
order of error in the determination of lattice parameters. (B) Grain size 
with annealing temperature. Bars indicate the ${\rm \pm1.5~nm}$ error in
determination of grain size. These results were obtained from the 
X-Ray Diffraction Patterns.}
\end{figure}
\newpage
\begin{figure}[h!]
\begin{center}
\epsfig{file=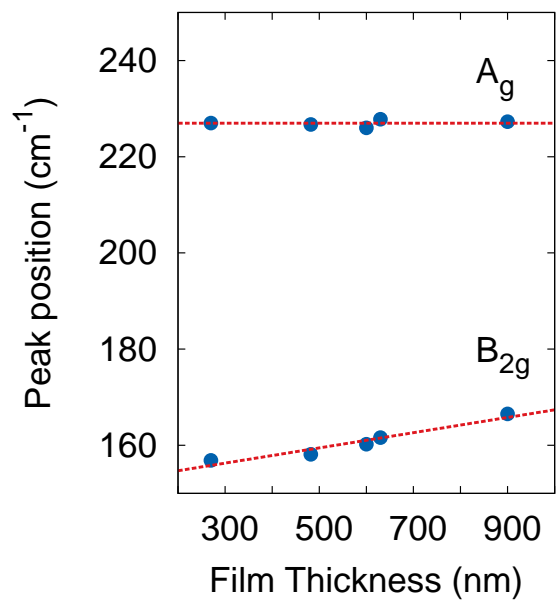, width=3.25in, angle=-90}
\end{center}
\label{raman2}
\caption{\sl The variation in Raman ${\rm A_g}$ and ${\rm B_{2g}}$ peak 
positions of annealed (373~K and 473~K) SnS films with film thickness.}
\end{figure}
\newpage
\begin{figure}[h!]
\begin{center}
\epsfig{file=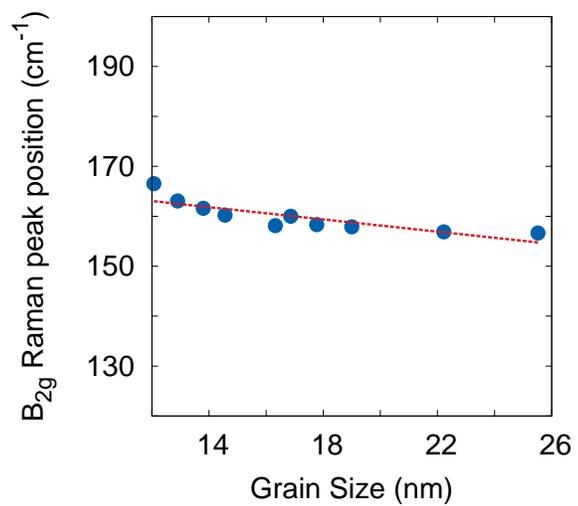, width=2.75in, angle=-90}
\end{center}
\caption{\sl The ${\rm B_{2g}}$ peak position of Raman spectra varies with
grain size. The graph shows shift of Raman peak to lower energy level with
increasing grain size. }
\label{result}
\end{figure}
\newpage
\begin{figure}[h!]
\begin{center}
\epsfig{file=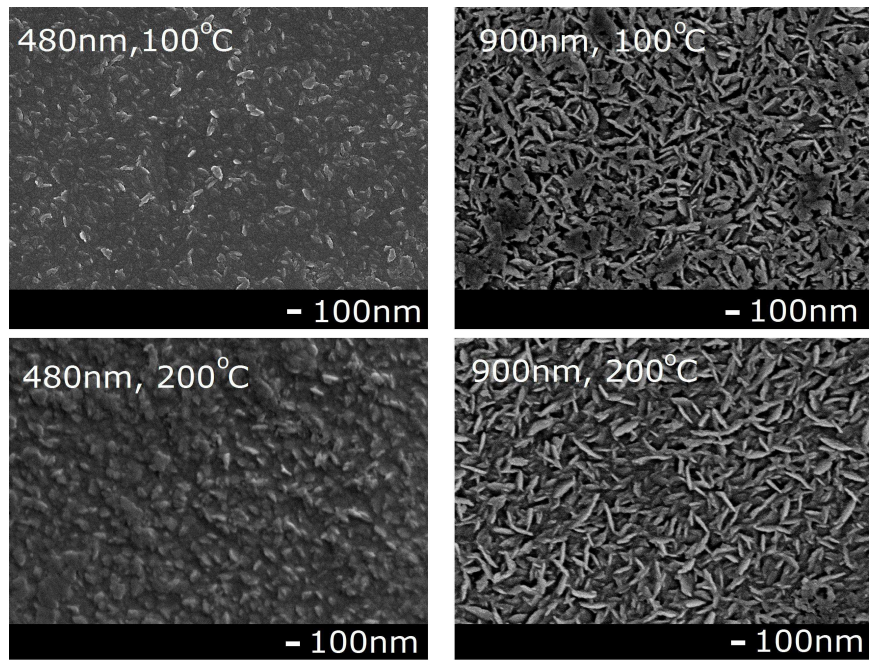, width=4.5in, angle=-0}
\end{center}
\label{fig4}
\caption{\sl SEM micrographs show grain size improvement in thinner film
(480~nm) when annealed at higher temperature. However, the 900~nm films show
no variation in grain size on annealing.}
\end{figure}
\newpage
\begin{figure}[h!]
\begin{center}
\epsfig{file=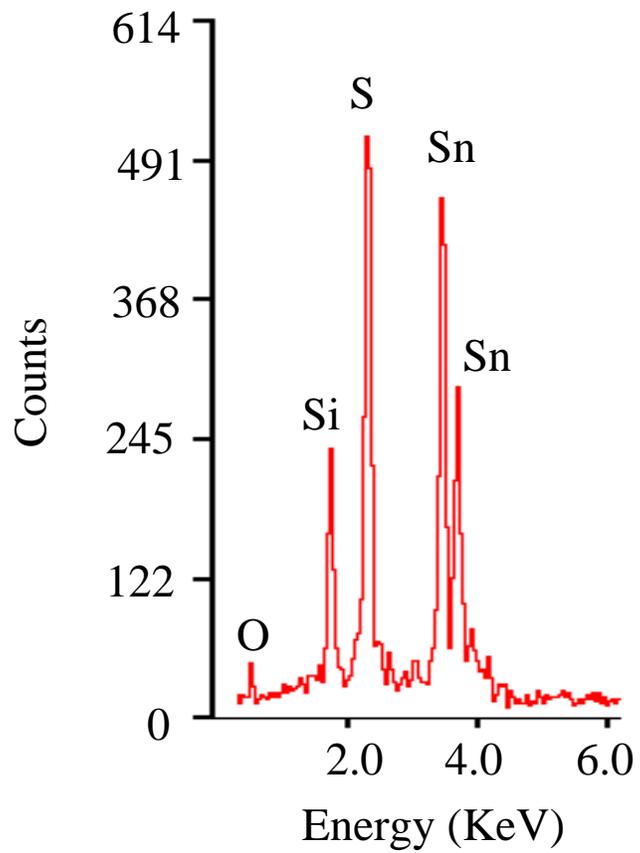, width=3.75in, angle=-0}
\end{center}
\label{fig4}
\caption{\sl Energy Dispersive X-Ray (EDX) Spectroscopy of a larger area of
900~nm thick film, vacuum annealed at 473~K, show no oxidation of the
sample. The minimal Silicon and Oxygen present are from the substrate.}
\end{figure}
\newpage

\begin{figure}[h!]
\begin{center}
\epsfig{file=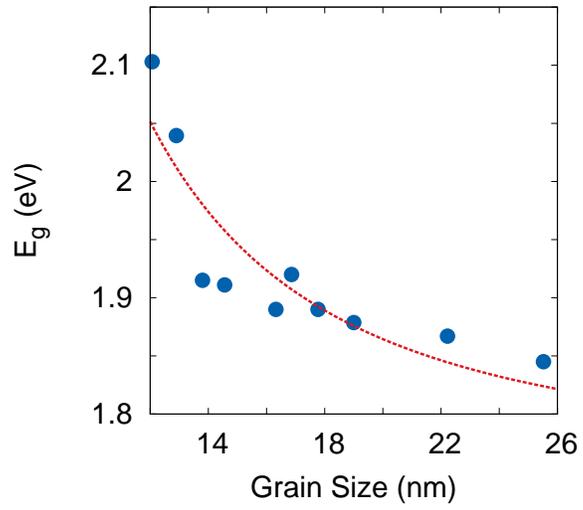, width=2.75in, angle=-90}
\end{center}
\caption{\sl The band-gap of annealed SnS films were found to decrease with
increasing grain size. The curve fit line agrees with theoretical model
that explains variation of band-gap with grain size.}
\label{result1}
\end{figure}

\end{document}